\documentclass[useAMS]{mn2e}

\usepackage{times}
\usepackage{graphicx}
\usepackage{amsmath}
\usepackage{amssymb}
\usepackage{color}
\newcommand{\beq}{\begin{equation}}
\newcommand{\bea}{\begin{eqnarray}}
\newcommand{\eeq}{\end{equation}}
\newcommand{\eea}{\end{eqnarray}}

\title[GRB afterglows in magnetized stellar winds]{Gamma-ray bursts
  afterglows in magnetized stellar winds}

\author[M. Lemoine and G. Pelletier]
{Martin Lemoine$^{1}$\thanks{e-mail:{\tt lemoine@iap.fr}} and
Guy Pelletier$^{2}$\thanks{e-mail:{\tt
     guy.pelletier@obs.ujf-grenoble.fr}}\\
$^{1}$ Institut d'Astrophysique de Paris, \\ 
	CNRS, UPMC, 
	98 bis boulevard Arago, F-75014 Paris, France\\
$^{2}$ Laboratoire d'Astrophysique de Grenoble, \\
	CNRS, Universit\'e Joseph Fourier II,
	BP 53, F-38041 Grenoble, France; \\
}

\begin{document}

\date{}

\pubyear{2008}

\maketitle

\label{firstpage}

\begin{abstract}
  Recent analytical and numerical work argue that successful
  relativistic Fermi acceleration requires a weak magnetization of the
  unshocked plasma, all the more so at high Lorentz factors. The
  present paper tests this conclusion by computing the afterglow of a
  gamma-ray burst outflow propagating in a magnetized stellar wind
  using ``ab initio'' principles regarding the microphysics of
  relativistic Fermi acceleration. It is shown that in magnetized
  environments, one expects a drop-out in the X-ray band on sub-day
  scales as the synchrotron emission of the shock heated electrons
  exits the frequency band. At later times, Fermi acceleration becomes
  operative when the blast Lorentz factor drops below a certain
  critical value, leading to the recovery of the standard afterglow
  light curve. Interestingly, the observed drop-out bears resemblance
  with the fast decay found in gamma-ray bursts early X-ray
  afterglows.
\end{abstract}

\begin{keywords} shock waves -- acceleration of particles --
gamma-ray bursts
\end{keywords}

\section{Introduction}
The prompt emission of gamma-ray bursts (GRB) is followed by an
afterglow phase commonly attributed to the synchrotron emission of
shock accelerated electrons (M\'esz\'aros \& Rees 1997). As the blast
wave sweeps up matter and decelerates, the dissipated power decreases
and the emission shifts to longer wavebands (e.g., Piran 2005). To
model this afterglow emission, one usually encodes the acceleration
physics in a minimal/maximal Lorentz factor ($\gamma_{\rm
  min}\,/\,\gamma_{\rm max}$), in the spectral index $s$ of the
electron spectrum, in the fraction $\epsilon_e$ of the dissipated
energy that is carried by these electrons and in the fraction
$\epsilon_B$ stored in magnetic turbulence.

However, our understanding of relativistic Fermi acceleration has made
significant progress in the last decade, to an extent that motivates a
direct test against observational data. The convergence of analytical
calculations and extensive particle-in-cell (PIC) numerical
calculations has led in particular to the following picture. At
(superluminal) ultra-relativistic shock waves, Fermi power-laws cannot
develop because the particles get advected to the far downstream along
with the magnetic field lines to which they are tied (Begelman \& Kirk
1990), unless strong turbulence has been excited on scales
significantly smaller than their Larmor radius (Lemoine et al. 2006;
Niemiec et al. 2006; Pelletier et al. 2009). In very weakly magnetized
shocks, such turbulence can be excited by micro-instabilities in the
shock precursor and therefore Fermi acceleration can develop, as
confirmed by recent PIC simulations (Sironi \& Spitkovsky 2011). The
critical level of magnetization below which this turbulence develops
depends on the shock Lorentz factor (Lemoine \& Pelletier 2010, 2011)
as indeed, such instabilities can grow only if their growth timescale
is shorter than the timescale on which the unshocked plasma crosses
the shock precursor and, the stronger the upstream background
magnetization, or the larger the shock Lorentz factor, the shorter the
precursor.

In practice, one may expect Fermi acceleration to proceed unhampered
if the blast wave propagates in a weakly magnetized external medium
such as the interstellar medium (ISM). In magnetized stellar winds,
however, one might expect to see signatures of the above microphysics
of Fermi acceleration, all the more so at early stages when the blast
Lorentz factor is large. Such signatures would open a window on the
physics of collisionless relativistic shocks as well as on the
astrophysics of GRB afterglows.  This motivates the present study,
which proposes to compute the afterglow light curve of a gamma-ray
burst propagating in a magnetized stellar wind from ``ab initio''
principles regarding Fermi acceleration.

The recent studies of Li \& Waxman (2006) and Li (2010) offer an
interesting perspective on this problem. From the observation of X-ray
afterglows on sub-day scales, these authors infer a strong lower bound
on the upstream magnetic field of gamma-ray bursts afterglows, $B_{\rm
  u}\gtrsim 200\,\mu{\rm G}\, n_{\rm 0}^{5/8}$ ($n_{0}$ the upstream
density in cm$^{-3}$); Li (2010) actually derives a significantly
stronger bound by considering on equal grounds the long lived high
energy emission $>100\,$MeV. This implies that either
micro-instabilities have grown and excited the magnetic field to the
above values, or the pre-existing magnetic field itself satisfies this
bound. While the former is expected if the circumburst medium is ISM
like, the latter corresponds to a magnetized circumburst medium. It is
this possibility that will be addressed and tested in the present
work.

Section~\ref{sec:model} presents an analytical discussion of the model
and is followed by numerical calculations of the light curve in
Sec.~\ref{sec:lcurve}. Section~\ref{sec:disc} discusses the results in
the light of modern X-ray afterglows.

\section{Physical model}\label{sec:model}
We consider the following fiducial values for the parameters
characterizing the afterglow. The ejecta is composed of a homogeneous
shell of width $\Delta =c T$ with $T=10T_{1}\,$s in the stationary
frame, with (isotropic equivalent) bulk kinetic energy
$E=10^{54}\,E_{54}\,$erg and Lorentz factor $\gamma_{\rm
  ej}=300\gamma_{\rm ej,2.5}$. This outflow impinges on a stellar wind
with density profile $\rho_{\rm w} = A r^{-2}$, with $A = 5\times
10^{11} A_{*}\,$g/cm and (toroidal) magnetic field profile $B_{\rm
  w}=10^3B_* \,{\rm G}\, \left(r/10^{12}\,{\rm cm}\right)^{-1}$; the
variables $A_*$ and $B_*$ encode our uncertainty on the density and
the magnetic field. The value $B_*=1$ corresponds to the formation of
a magnetar with surface field $\sim 10^{15}\,$G after the collapse of
a $10R_\odot$ progenitor star. The magnetic field of Wolf-Rayet stars,
which are considered as potential progenitor stars for gamma-ray
bursts, are not known, but surface values as large as $1-10\times
10^3\,$G have been considered (Ignace et al. 1998). For these
parameters, the magnetization $\sigma_{\rm w}\equiv B_{\rm
  w}^2/\left(4\pi \rho_{\rm w}c^2\right)\simeq 1.8\times 10^{-4}
B_*^2A_*^{-1}$; it does not depend on $r$. Note that the central
assumption here is that of a relatively high magnetization of the
external medium; the density profile does not play a crucial role and
similar effects can be observed in a constant density medium of
sufficient magnetization, as discussed briefly in Sec.~\ref{sec:disc}.

The proper density of the ejecta $n_{\rm ej} = E_{\rm ej}/\left(4\pi
  r^2 \gamma_{\rm ej}^2 \Delta m_pc^2\right)$; therefore the density
contrast between the ejecta and the external medium is not very large,
$\left(n_{\rm ej}/n_{\rm w}\right)^{1/2} \simeq 81
T_{1}^{-1/2}\gamma_{\rm ej,2.5}^{-1}E_{54}^{1/2}A_*^{-1/2}$, implying
that the reverse shock propagates at relativistic speeds in the ejecta
(Sari \& Piran 1995). The shocked material -- which we denote as the
blast -- thus moves with initial Lorentz factor
\begin{equation}
  \gamma_{\rm b} \simeq \frac{\gamma_{\rm ej}^{1/2}}{\sqrt{2}}
  \left(\frac{n_{\rm ej}}{n_{\rm w}}\right)^{1/4}
  \simeq 110\, T_{1}^{-1/4}E_{54}^{1/4}A_*^{-1/4}\ .\label{eq:gammab}
\end{equation}
It remains constant as long as the reverse shock is crossing the shell
(Sari \& Piran 1995, Beloborodov \& Uhm 2006).  Approximating the
velocity of the reverse shock as $c$ in the ejecta frame, the reverse
shock has crossed the outflow at radius $r_\times =\gamma_{\rm b}^2 c
T$ (in the stationary frame), corresponding to observer time
$t_{\times}\simeq 5\,{\rm sec}\,(1+z)T_{1}$, with $z$ the GRB
redshift.

Beyond $r_\times$, the blast Lorentz factor decreases according to
$\gamma_{\rm b}\simeq \gamma_{\rm
  b,\times}\left(r/r_\times\right)^{-1/2}$ in the adiabatic regime,
$\gamma_{\rm b,\times}$ corresponding to Eq.~(\ref{eq:gammab}). The
relationship between observer time, radius and blast Lorentz factor
then becomes $\left(t_{\rm obs}/t_\times\right) \simeq
\left(r/r_\times\right)^2 \simeq \left(\gamma_{\rm b}/\gamma_{\rm
    b,\times}\right)^{-4}$.

We now come to the modelling of the electron population in the
blast. Following Lemoine \& Pelletier (2010, 2011), we define the
parameter $Y_{\rm inst} \equiv \xi_{\rm b}^{-1}\sigma_{\rm
  w}\gamma_{\rm sh}^2$, which characterizes whether instabilities may
develop or not in the shock precursor, hence whether Fermi cycles can
develop or not. The parameter $\xi_{\rm b}$ denotes the fraction of
incoming matter energy through the shock $4\gamma_{\rm b}^2\rho_{\rm
  w}c^2$ that is carried by the accelerated and returning particles
(i.e. the beam). By returning, it is meant those incoming protons that
are reflected on the shock front, which constitute an essential
ingredient of the shock formation. These reflected protons exist even
in the absence of Fermi powerlaws. Through mixing with the unshocked
plasma, these returning particles (along with the accelerated
particles) induce two-stream or filamentation micro-instabilities in
the shock precursor, on scales close to the electron to ion skin depth
$c/\omega_{\rm pe}\rightarrow c/\omega_{\rm pi}$. The filamentation
instability has time to grow only if $Y_{\rm inst}\,\ll\,1$, while
other two stream instabilities may grow faster but are inhibited once
the background electrons are heated to relativistic temperatures in
the shock precursor (Lemoine \& Pelletier 2011). For this reason, we
consider only the growth of the filamentation instability in the
following.  We define a threshold value $Y_{\rm c}$ such that if
$Y_{\rm inst}<Y_{\rm c}$, micro-instabilities can grow and allow Fermi
cycles to develop, as discussed further below, while if $Y_{\rm
  inst}>Y_{\rm c}$, instabilities cannot grow, hence Fermi cycles do
not develop.

One must expect $Y_{\rm inst}>Y_{\rm c}$ in the early stages of the
afterglow, since 
\begin{equation}
  Y_{\rm inst}\simeq 43\,
  B_*^2A_*^{-3/2}T_{1}^{-1/2}E_{54}^{1/2}\xi_{\rm b,-1}^{-1}
\left(\frac{r}{r_\times}\right)^{\alpha_Y}\ ,
\end{equation}
with $\alpha_Y=0$ for $r<r_\times$ and $\alpha_Y=-1$ for
$r>r_\times$. PIC simulations indicate that $\xi_{\rm b,-1}\equiv
\xi_{\rm b}/0.1\simeq 1$ (Sironi \& Spitkovsky 2011).  

Early on, as $Y_{\rm inst}>Y_{\rm c}$, micro-instabilities are
quenched by advection of the plasma through the shock front, hence the
magnetic field is everywhere transverse to the shock normal without
substantial inhomogeneity on short scales. In this case, Fermi
acceleration cannot develop as particles are advected with the
magnetic field lines to the far downstream. Nevertheless, the
electrons acquire part of the kinetic energy of the incoming protons
in the shock transition (as viewed in the shock frame). A detailed
understanding of this process is still lacking but current PIC
simulations confirm the above, even in the absence of filamentation in
the precursor. In particular, Sironi \& Spitkovsky (2011) observe that
$\epsilon_e$ reaches the value of $0.1$ at a magnetization
$\sigma_{\rm u}=10^{-4}$, for $\gamma_{\rm sh}\simeq 20$ and
larger. We adopt this value in the following. For simplicity, we model
the shock heated electron distribution as a restricted powerlaw with
$\gamma_{\rm max}=3\gamma_{\rm min}$. The minimal Lorentz factor
$\gamma_{\rm min}$ is then related to $\epsilon_e$ through
$\gamma_{\rm min}= \epsilon_e \gamma_{\rm b} (m_p/m_e)a_{\rm s}$ with
$a_{\rm s}=\left[(s-2)/(s-1)\right] \left[1- (\gamma_{\rm
    max}/\gamma_{\rm min})^{1-s}\right] \left[1- (\gamma_{\rm
    max}/\gamma_{\rm min})^{2-s}\right]^{-1}$ a normalization
prefactor of order unity, which depends (slightly) on the modelling of
the energy distribution; we adopt $s=2.4$, an ad-hoc choice here as
well motivated by simplicity (i.e. $s$ will not change once Fermi
acceleration becomes effective).  Although the electrons are heated in
the shock transition, the magnetic field is only compressed, so that
the magnetic field in the blast frame $B_{\rm b} = 4\gamma_{\rm b}
B_{\rm w}$.  In terms of the conventional parameter $\epsilon_B$
describing the fraction of energy carried by the magnetic field in the
blast, $\epsilon_B=B_{\rm b}^2/(8\pi e_{\rm b}) = 2\sigma_{\rm w}$.

As the blast Lorentz factors decreases beyond $r_\times$, so does
$Y_{\rm inst}$, until $Y_{\rm inst}<Y_{\rm c}$ eventually. The
filamentation instability now has several to many $e-$folds of growth
times before the plasma is advected through the shock front. This has
several consequences of importance. First of all, the upstream
electrons are heated in the micro-turbulence in the shock precursor
(Spitkovsky 2008, Lemoine \& Pelletier 2011) and they therefore reach
rough equipartition with the incoming protons after the shock
transition, as observed in PIC simulations (Sironi \& Spitkovsky
2011). This implies $\epsilon_e\simeq 0.5$. Furthermore, a
micro-turbulent magnetic field is generated on skin depth scales up to
$\epsilon_B$ of a few percents. We adopt $\epsilon_B=0.05$ as a
fiducial value in what follows, in the absence of more detailed
results from PIC simulations in the parameter range of
interest. Finally, as discussed above, the micro-turbulence unlocks
the particles off the magnetic field lines and allow them to scatter
repeatedly back and forth the shock wave, leading to a power-law
extension beyond the (relativistic) thermal population. This fact has
been clearly observed in the PIC simulations of Sironi \& Spitkovsky
(2011), for $\gamma_{\rm sh}\simeq 20$ and upstream magnetization
$10^{-5}$ (and mass ratio $m_p/m_e=16$). Note that the same
simulations at magnetization $10^{-4}$ do not observe signs of Fermi
acceleration, suggesting that $Y_{\rm c}\lesssim 0.5$. In the
following, we keep manifest the dependence on $Y_{\rm c}$. To model
the resulting electron distribution, we use a power-law between
$\gamma_{\rm min}$ and $\gamma_{\rm max}$, with $\gamma_{\rm min}$
related to $\epsilon_e$ as before (although here,
$\epsilon_e\simeq0.5$); we keep $s=2.4$. This implies that we do not
distinguish between the thermal and the powerlaw tail populations;
this is a good approximation, since both radiate in synchrotron, hence
the above simplification only affects the flux normalization at high
energies.

Synchrotron energy losses provide an upper bound on the maximal
Lorentz factor, $\gamma_{\rm max}\,\lesssim\,4\times 10^6 E_{54}^{1/8}
T_{1}^{3/8} A_*^{-3/8}\epsilon_{B,-1.3}^{-1/4} (r/r_\times)^{3/4}$ (at
$r>r_\times$, with $\epsilon_{B,-1.3}=\epsilon_B/0.05$). In the
present case, the actual limiting factor for $\gamma_{\rm max}$ comes
from the scattering properties of accelerated particles in the
micro-turbulence, as discussed in Pelletier et al. (2009). Indeed,
Fermi cycles can develop if the angular scattering in the
micro-turbulence dominates over the regular Larmor orbits in the
background field on a cycle timescale, which requires $r_{\rm L,0}
\lesssim \ell_{\rm c} (\delta B/B)^2$, with $r_{\rm L,0}$ the Larmor
radius in the background field $4\gamma_{\rm b}B_{\rm w}$, $\ell_{\rm
  c}$ the micro-turbulence scale and $\delta B$ the microturbulence
strength. Both simulations (e.g., Sironi \& Spitkovsky 2011) and
analytical arguments (e.g. Lemoine \& Pelletier 2011) indicate that
the relevant length scale is the ion skin depth $\delta_{\rm i}$ (as
measured in the upstream frame), while $(\delta B/B)^2 \simeq
\epsilon_B/\sigma_{\rm w}$. This implies a maximal Lorentz factor
\begin{equation}
  \gamma_{\rm max}\,\sim\, 9\times 10^5
  E_{54}^{1/4}T_{1}^{-1/4}A_*^{1/4}\epsilon_{B,-1.3}B_*^{-1}\left(\frac{r}{r_\times}\right)^{-1/2}\  ,
\label{eq:gmax-deltaB}
\end{equation}
assuming here $r>r_\times$.

Several remarks are in order at this stage. We do not consider the
issue of the evolution of the micro-turbulence in the downstream,
which remains an open problem in this field (Gruzinov \& Waxman 1999,
Medvedev \& Loeb 1999). The typical Larmor radius of electrons of
Lorentz factor $\gamma_{\rm b}m_p/m_e \sim \gamma_{\rm min}$ reads
$r_{L} \sim \delta_{\rm i}/\epsilon_B^{1/2}$, hence the first
generations of accelerated electrons only probe the vicinity of the
shock front in terms of $\delta_{\rm i}$, where the turbulence should
not have evolved strongly. In our case, the electron population
develops on a a dynamic range $\sim \delta B/B \sim
(\epsilon_B/\sigma_{\rm w})^{1/2}\lesssim 10^2$, see
Eq.~(\ref{eq:gmax-deltaB}). Therefore the highest energy electrons
explore the blast up to $\sim 10^4\delta_{\rm i}$ (given that the
scattering length in the micro-turbulence scales as $\gamma_e^2$);
admittedly, one cannot exclude that the turbulence evolves on such
length scales. For reference, the ion skin depth $\delta_{\rm
  i}\simeq2.8\times 10^5\,{\rm
  cm}\,E_{54}^{1/2}T_{1}^{1/2}A_*^{-1}(r/r_\times)$.

We also neglect the influence of extra large scale sources of
turbulence, associated with e.g. instabilities of the blast itself
(e.g. Levinson 2010), or with the interactions of the shock with
inhomogeneities of the wind (e.g. Sironi \& Goodman 2007). This is
justified insofar as the strong background magnetic field effectively
prevents particles located further than $\sim r_{\rm L,0}$ away from
the shock front to return to the shock front, and $r_{\rm L,0}\sim
\delta_{\rm i}/\sigma_{\rm w}^{1/2}$ is already much smaller than the
typical scales at which such instabilities develop. This means that
particles that undergo Fermi cycles cannot experience turbulence
sourced on scales larger than $r_{\rm L,0}$. Finally, the present
study does not discuss the impact of pair loading in front of the
shock wave (e.g. Beloborodov 2005, Ramirez-Ruiz et al. 2007), which
will be addressed in a future work.

\section{Light curve}\label{sec:lcurve}
The above description provides the necessary ingredients to compute
the light curve. We will be mostly interested in the X-ray afterglow,
which probes the highest energy electron population at the early
stages of the afterglow. We rely on the model introduced by
Beloborodov (2005), which assumes that electrons cross the shock
front, get instantaneously accelerated to a powerlaw, then cool
adiabatically and through synchrotron / inverse Compton losses. This
model fits nicely the present description and the present hierarchy of
timescales: $t_{\rm acc}\,\ll\,t_{\rm loss}\,\ll\,\Delta\gamma_{\rm
  b}/c$ and $r_{\rm L,0}\,\ll\,\Delta\gamma_{\rm b}$ (the blast width
in the blast rest frame). We have added the spectral contribution of
fast cooling electrons to the model of Beloborodov (2005) in order to
discuss the X-ray light curve. We also take into account inverse
Compton cooling following the parametrization of Sari \& Esin (2001)
as discussed in Li \& Waxman (2006): in particular, at early times
when the blast magnetization $\epsilon_B=2\sigma_{\rm
  w}\,\ll\,\epsilon_e$ and electrons cool rapidly, the Compton
parameter $Y_{\rm IC}\simeq (\epsilon_e/\epsilon_B)^{1/2}$, while at
late times, in the slow cooling regime, $Y_{\rm IC}\sim 1$. The
radiative loss time is then written as $t_{\rm syn}/(1+Y_{\rm IC})$,
with $t_{\rm syn}$ the synchrotron loss time in the blast frame. At
$r_\times$ and beyond, Klein-Nishina effects are not significant since
$h\nu_{\rm min}\gamma_{\rm min}\sim \gamma_{\rm b}m_e c^2$ for the
fiducial values (at $r_\times$) and $\gamma_{\rm max}\sim 3\gamma_{\rm
  min}$ before recovery. The deceleration of the blast wave is
followed by solving the equations of the mechanical model of
Beloborodov \& Uhm (2006). The ejecta and the blast are assumed
homogeneous and once the reverse shock has crossed the ejecta, its
contribution is discarded from the equations of motion. We also assume
an adiabatic evolution of the blast wave. This is clearly justified at
early times, when $\epsilon_e=0.1$; at late times, $\epsilon_e=0.5$
but the emission takes place mostly in the slow cooling regime,
therefore this remains a reasonable approximation.

\begin{figure}
\includegraphics[width=0.49\textwidth]{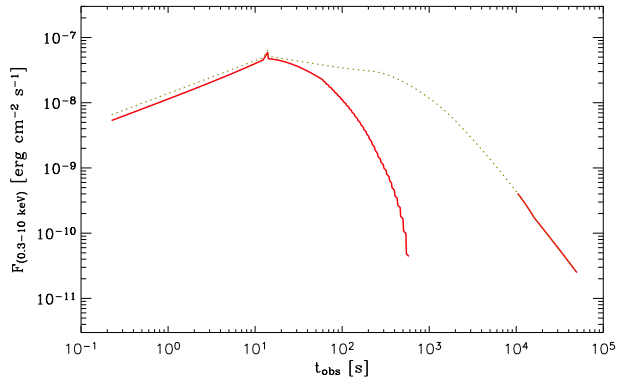}
\caption{X-ray light curve of a gamma-ray burst located at $z=1$,
  with fiducial parameters as described at the beginning of
  Sec.~\ref{sec:model}; in solid line, result of the model, revealing
  the flux drop-out at $t_{\rm obs}\sim100\,$s and the recovery at
  late times $\sim10^4\,$s. In dotted line, the same model, assuming
  however that micro-instabilities can grow at all times in front of
  the forward shock (thus implying $\epsilon_e=0.5$, $\epsilon_B=0.05$
  at all times); this situation also corresponds to what would be seen
  for magnetizations $\sigma_{\rm u}\,\lesssim\,10^{-6}$, all other
  parameters remaining unchanged.\label{fig:lcurve} }
\end{figure}
Figure~\ref{fig:lcurve} presents the resulting light curve in the
energy interval $0.3-10\,$keV. The parameters correspond to the
previous fiducial values; we have also adopted $\epsilon_B=0.05$,
$Y_{\rm c}=1$ and $z=1$. At early times, $t_{\rm obs}\,\ll\, 10^4\,$s,
$Y_{\rm inst}>Y_{\rm c}$ hence there is no Fermi power-law, only a
thermal electron population extending over half an order of magnitude,
implying that the synchrotron emission extends over an order of
magnitude. The (observer frame) frequency $\nu_{\rm min}$ associated
to $\gamma_{\rm min}$ reads
\begin{eqnarray}
  \nu_{\rm min}&\simeq& \frac{0.2}{1+z}\frac{e B_{\rm b}}{m_e
    c}\gamma_{\rm b}\gamma_{\rm min}^2\nonumber\\
  &\simeq&\frac{1.6\times 10^{18}\,{\rm
      Hz}}{1+z}\,\,E_{54}^{1/2}A_*^{-1/2}T_{1}^{-3/2}B_*\epsilon_{e,-1}^2
  \left(\frac{r}{r_\times}\right)^{\alpha_{\nu}}\ ,\label{eq:numin}
\end{eqnarray}
with $\alpha_{\nu}=-1$ for $r<r_\times$, $\alpha_{\nu}=-3$
otherwise. Consequently, for $r>r_\times$, meaning $t_{\rm
  obs}>5\,{\rm s}\,(1+z)T_{1}$, the minimum frequency drops rapidly
out of the X-ray band. This is accompanied by a drastic reduction in
flux as the maximal frequency $\nu_{\rm max}$ ($\sim 9\nu_{\rm min}$
in the absence of Fermi powerlaw) also exits progressively out the
X-ray domain. Given the strong dependence of $\nu_{\rm min}$ on $r$,
the drop-out occurs shortly after $t_\times$: in detail, defining the
drop-out time $t_{\rm d-o}$ as that at which $\nu_{\rm min}=0.7\times
10^{17}\,$Hz,
\begin{equation}
  t_{\rm d-o}\,\simeq\,110\,{\rm s}\,
  E_{54}^{1/3}A_*^{-1/3}B_*^{2/3}\epsilon_{e,-1}^{4/3}(1+z)^{1/3}\ .\label{eq:tdo}
\end{equation}
This timescale does not depend on the duration of the prompt emission
(although it cannot of course be shorter). The shape of the light
curve during the drop-out is affected by our assumption of a
restricted powerlaw; a more detailed modelling of the electron
spectral distribution (e.g. Giannios \& Spitkovsky 2009) is required
to refine the prediction for the light curve in this region. One
should also account for the delay associated with emission away from
the ligne of sight or from within the blast, which would lead to a
smoothing of the light curve on a timescale $\sim t_{\rm obs}$ at
$t_{\rm obs}$.

\begin{figure}
\includegraphics[width=0.49\textwidth]{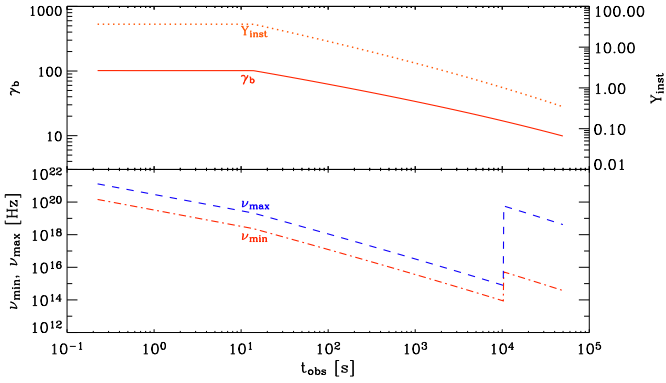}
\caption{Upper panel: evolution in time of the blast Lorentz factor
  (left $y-$axis) and of the instability parameter $Y_{\rm inst}$
  (right $y-$axis); the onset of Fermi acceleration occurs when
  $Y_{\rm inst}<1$ in the model shown in Fig.~\ref{fig:lcurve},
  corresponding to $t_{\rm obs}\simeq 10^4\,$s. Lower panel: evolution
  with time of the minimal and maximal observer frame electron
  synchrotron frequencies; at times $<10^4\,$s, Fermi acceleration
  does not take place, a powerlaw cannot develop hence $\nu_{\rm max}$
  is $\sim1$ order of magnitude larger than the frequency $\nu_{\rm
    min}$. Note that $\nu_{\rm min}$ exits the X-ray band around
  $t_{\rm obs}\sim 100\,$s. At late times, Fermi powerlaws develop and
  the range $\nu_{\rm min}-\nu_{\rm max}$ broadens significantly,
  leading to the recovery of the standard
  light curve.\label{fig:aftg-par} }
\end{figure}
The above simplified model predicts no flux in the X-ray band between
the completion of the drop-out, roughly a factor of a few beyond
$t_{\rm d-o}$ and the recovery, i.e. the time at which $Y_{\rm
  inst}=Y_{\rm c}$. This latter timescale $t_{\rm rec}$ can be written
\begin{equation}
  t_{\rm rec}\,\simeq\,0.9\times 10^4\,{\rm s}\, E_{54}A_*^{-3}B_*^4Y_{\rm
    c}^{-2}\xi_{\rm b,-1}^{-2}(1+z)\ .\label{eq:trec}
\end{equation}
At $t_{\rm rec}$, Fermi cycles develop on a very short timescale
compared to the dynamical timescale, hence emission can take place up
the maximal frequency $\nu_{\rm max}$ corresponding to $\gamma_{\rm
  max}$, with
\begin{equation}
  \nu_{\rm max}\,\simeq\, 7.6\times 10^{18}\,{\rm Hz}\, E_{54}^{-1} A_*^{11/2}B_*^{-8} 
  Y_{\rm c}^3 \epsilon_{B,-1.3}^{5/2} \xi_{\rm b,-1}^3
  \left(\frac{t}{t_{\rm rec}}\right)^{-3/2}\ .
\end{equation}
The strong dependence of this maximal frequency on the parameters
suggests that a variety of effects could take place; in particular,
one might observe a weak recovery in the X-ray band or even no
recovery at all. Caution has to be exerted however, since when $t_{\rm
  obs}\gtrsim 10^4\,$s the Lorentz factor of the blast has dropped to
moderate values $\sim 20$, hence additional effects may come into
play.  In particular, one cannot rule out the emergence of new
instabilities at scales larger than $\delta_{\rm i}$ that would push
$\gamma_{\rm max}$ hence $\nu_{\rm max}$ to much larger values.  At
even later times, jet sideways expansion affects the dynamical
evolution; as the above one dimensional model ignores such effects, we
stop the calculation at $t=10^5\,$s.

Figure~\ref{fig:aftg-par} summarizes the evolution with $t_{\rm obs}$
of the main parameters and allows to understand better the behaviour
of the X-ray light curve shown in Fig.~\ref{fig:lcurve}. Note that one
does not expect a drop-out in the optical, since $\nu_{\rm
  max}\gtrsim10^{15}\,$Hz for the present fiducial values and
$\nu_{\rm min}$ crosses the optical shortly before $t_{\rm rec}$. Such
a drop-out could only be seen if $t_{\rm rec}$ were made much larger
than $10^4\,$s, e.g. by increasing the magnetization, see
Eqs.~(\ref{eq:numin}),(\ref{eq:tdo}),(\ref{eq:trec}).

\section{Discussion}\label{sec:disc}
Using ``ab initio'' principles of relativistic Fermi acceleration now
tested in extensive PIC shock simulations, we have calculated the
X-ray afterglow light curve of a GRB propagating in a magnetized
stellar wind of magnetization $\sigma_{\rm u}\sim 10^{-4}$, assuming
otherwise standard GRB parameters. We have shown that the inhibition
of relativistic Fermi acceleration in magnetized shocks at high
Lorentz factor leaves a distinct signature in the light curve, in the
form of a fast drop-out shortly after the end of the prompt emission,
around $t_{\rm obs}\sim 100\,$s, with a recovery at late times $\sim
10^4\,$s. The latter depends more strongly on the model parameters, in
particular magnetization so that one may envisage a variety of
situations beyond that described: e.g., no drop-out if $\sigma_{\rm
  u}\lesssim 10^{-6}$ -- ceteris paribus -- or a drop-out with no
recovery if $\sigma_{\rm u}\gtrsim 10^{-3}$. Although we have
calculated the light curve for a stellar wind profile, similar effects
might be observed for a constant density circumburst medium, provided
the magnetization is large enough. In particular, one would observe a
drop-out at $t_{\rm d-o}\simeq 180\,{\rm sec}\,(1. +z)^{1/3}
B_{-3}^{2/3}E_{54}^{1/3}n_0^{-1/3} \epsilon_{e,-1}^{4/3}$ for a
magnetic field $B=10^{-3}B_{-3}\,$G and density $n=n_0\,$cm$^{-3}$,
followed by recovery at $t_{\rm rec}\simeq 10^4\,{\rm sec}\, (1. +z)
B_{-3}^{8/3} E_{54}^{1/3}n_0^{-5/3} Y_{\rm c}^{-4/3} \xi _{{\rm
    b},-1}^{-4/3}$. Therefore the present results extend beyond the
stellar wind case and may be applicable to both long and short GRBs.

Interestingly, recent Swift observations have revealed a rather
complex early X-ray afterglow light curve in a subset of GRBs, with a
fast decay at $t_{\rm obs}\sim100\,$s followed by a form of plateau
that joins a more standard light curve at later times $\gtrsim
10^4\,$s (Nousek et al. 2006, O'Brien et al. 2006). High latitude
emission is considered as a possible explanation for the steep decay
phase, although modelling the plateau phase with the afterglow brings
in additional constraints on the overall GRB model (e.g. Panaitescu
2007).  The present scenario could account for two of these observed
features -- the initial fast decay and the late time recovery -- but
it does not explain the emergence of the plateau. The following
briefly addresses these issues in turn.

Regarding the fast decay phase, the present scenario predicts an
exponential decay and a clear spectral transition from hard to soft as
the peak of the emission exits the X-ray band. One does not therefore
expect a perfectly smooth transition from the prompt emission to the
fast decay phase.  As discussed in Sec.~\ref{sec:lcurve}, additional
theoretical developments are required to provide a detailed light
curve around $100\,$s. Nevertheless, it is of interest to note that
Sakamoto et al. (2007) have reported evidence for an exponential decay
component on top of a powerlaw decaying component in the early X-ray
light curve. Furthermore Zhang et al. (2007) have observed a
pronounced hard-to-soft spectral transition during the fast decay in
two thirds of GRBs that show a fast decay (see also Yonetoku et
al. 2008). Their phenomenological model involves an exponentially
cut-off energy spectrum, the peak energy $E_{\rm c}$ of which moves
out of the X-ray band during the fast decay according to $E_{\rm
  c}\propto t_{\rm obs}^{-\alpha_{\rm c}}$ with $\alpha_{\rm c}\simeq
1-1.5$. This fits quite well the present picture, considering in
particular that $\nu_{\rm min}\propto t_{\rm obs}^{-1.5}$
[Eq.~(\ref{eq:numin})]. We also note that some short GRBs show an
exponentially decaying light curve around 100s, well beyond the prompt
emission, accompanied by spectral evolution, such as GRB050724
(Campana et al. 2006), while some show fast decay without apparent
late time recovery (e.g. GRB051210, GRB060801, see Nakar 2007).

Regarding the shallow decay phase, a contribution from the reverse
shock has been envisaged in Uhm \& Beloborodov (2007) and Genet et
al. (2007), although some tuning appear required to ensure a smooth
transition to the recovery phase. After the present work was
submitted, a paper by Petropoulou et al. (2011) appeared, arguing that
the shallow decay phase can be accounted for by the low energy tail of
the synchrotron self Compton component. Alternatively, one could try
to explain the shallow decay phase with an inefficient contribution of
the forward shock, the efficiency increasing with time and reaching
its maximum at recovery of the standard light curve. This could be
accomplished if a small fraction $\chi\epsilon_e$ with $\chi\ll1$ is
stored in an accelerated electron powerlaw at that time -- beyond the
thermal component that amounts to $\epsilon_e(1-\chi)\sim\epsilon_e$
-- with $\chi$ rising up to $\sim1$ at recovery. Granot et al. (2006)
and Ioka et al. (2006) have proposed a similar scenario, with varying
microphysical parameters during the shallow decay phase. Our model
assumes a sharp transition at $Y_{\rm inst}=1$ between no acceleration
(i.e. $\chi=0$) and fully efficient acceleration ($\chi=1$), but what
actually happens at $Y_{\rm inst}\sim {\cal O}(1)$ when a few efolds
of growth of the turbulence can occur is not known, since we have at
our disposal the results of only two simulations at $\sigma_{\rm
  u}=10^{-4}$ and $10^{-5}$. Moreover, one should recall that current
PIC simulations probe tiny timescales in regards of the GRB timescales
and that these simulations do not yet converge to a stationary shock
state (Keshet et al. 2009). One thus cannot exclude that inefficient
acceleration occurs at $Y_{\rm inst}\sim {\cal O}(1)$ but goes
undetected in current simulations; dedicated PIC simulations on long
timescales appear required to probe this transition region.
Alternatively, if the jet is structured in energy and Lorentz factor
per solid angle, an observer may receive emission from regions of
different Lorentz factors than that on the line of sight
(e.g. Panaitescu 2007); if the Lorentz factors in those off-axis
regions are such that $Y_{\rm inst}<Y_{\rm c}$, one might detect low
flux emission, corresponding to $\chi<1$ and possibly a shallow decay
phase. Yet another possibility is that of a clumpy circumburst medium,
with clumps of various sizes, provided $r_{\rm c}\,\ll\, r/\gamma_{\rm
  b}$~\footnote{clumps at the base of the wind are indeed expected to
  have a radius $\lesssim 0.01r$ (e.g. Owocki 2011) and $\gamma_{\rm
    b}$ decreases with increasing $r$}. As the causal region of
lateral extent $r/\gamma_{\rm b}$ contains many clumps, one does not
expect a bumpy signature in the light curve. However, one would
collect only a fraction $\chi<1$ of the expected X-ray flux due to
acceleration in the fraction $f_{\rm c}<1$ of the clumps that carry a
magnetization such that $Y_{\rm inst}<Y_{\rm c}$ at a given time. As
the overall density and Lorentz factor decrease, $f_{\rm c}$ increases
and so does $\chi$ until recovery, which corresponds to $Y_{\rm
  inst}\lesssim Y_{\rm c}$ in the smallest scale clumps that carry
most of the mass. In each of the above scenarios, one would expect a
smooth transition in the light curve with no spectral evolution
between the shallow decay phase and the late time normal decay phase,
as reported by Nousek et al. (2006) and O'Brien et al. (2006).

More work is certainly warranted to discuss these aspects in more
detail and to compare the properties of the light curve to
observational data in the relevant wavelength domains. One may in
particular expect the inverse Compton GeV emission to provide further
constraints on the present scenario.
\smallskip

{\bf Acknowledgments:} we acknowledge support from the CNRS PEPS/PTI
Program and from the GDR PCHE.


\begin{thebibliography}{}

\bibitem[]{} Beloborodov, A., 2005, ApJ, 627, 346

\bibitem[]{} Beloborodov, A., Uhm, L., 2006, ApJ, 651, L1

\bibitem[]{} Begelman, M. C., Kirk, J. G., 1990, ApJ, 353, 66

\bibitem[]{} Campana, S., Tagliaferri, G., Lazzati, D. et al., 2006,
  AA, 454, 113

\bibitem[]{} Genet, F., Daigne, F., Mochkovitch, R., 2007, MNRAS, 381, 732

\bibitem[]{} Giannios, D., Spitkovsky, A., 2009, MNRAS, 400, 330

\bibitem[]{} Granot, J., K\"onigl, A., Piran, T., 2006, MNRAS, 370,
  1946

\bibitem[]{} Gruzinov, A., Waxman, E., 1999, ApJ, 511, 852

\bibitem[]{} Ignace, R., Cassinelli, J. P., Bjorkman, J. E., 1998,
  ApJ, 505, 910

\bibitem[]{} Ioka, K., Toma, K., Yamazaki, R., Nakamura, T., 2006, AA,
  458, 7

\bibitem[]{} Keshet, U., Katz, B., Spitkovsky, A., Waxman, E., 2009,
  ApJ, 693, L127

\bibitem[]{} Lemoine, M., Pelletier, G., Revenu, B., 2006, ApJ, 645, L129

\bibitem[]{} Lemoine, M., Pelletier, G., 2010, MNRAS, 402, 321

\bibitem[]{} Lemoine, M., Pelletier, G., 2011, arXiv:1102.1308

\bibitem[]{} Levinson, A., 2010, ApJ, 705, L213

\bibitem[]{} Li, Z., Waxman, E., 2006, ApJ, 651, L328

\bibitem[]{} Li, Z., 2010, arXiv:1004.0791

\bibitem[]{} Medvedev, M. V., Loeb, A., 1999, ApJ, 526, 697

\bibitem[]{} M\'esz\'aros, P., Rees, M., 1997,  ApJ, 476, 232

\bibitem[]{} Nakar, E., 2007, Phys. Rep., 442, 166

\bibitem[]{} Niemiec, J., Ostrowski, M., Pohl, M., 2006, ApJ, 650, 1020

\bibitem[]{} Nousek, J. A., Kouveliotou, C., Grupe, D. et al., 2006, ApJ,
  642, 389

\bibitem[]{} O'Brien, P. T., Willingale, R., Osborne, J. et al., 2006,
  ApJ, 647, 1213

\bibitem[]{} Owocki, S., 2011, Bull. Soc. Roy. Sc. Li\`ege, vol. 80, p. 16

\bibitem[]{} Panaitescu, A., 2007, MNRAS, 379, 331

\bibitem[]{} Pelletier, G., Lemoine, M., Marcowith, A., 2009, MNRAS,
  393, 587

\bibitem[]{} Petropoulou, M., Mastichiadis, A., Piran, T., 2011, AA,
  531, 76

\bibitem[]{} Piran, T., 2005,  Rev. Mod. Phys., 76, 1143

\bibitem[]{} Ramirez-Ruiz, E., Nishikawa, K.-I., Hededal, C. B., 2007,
  ApJ 671, 1877

\bibitem[]{} Sakamoto, T., Hill, J. E., Yamazaki, R. et al., 2007, ApJ
  669, 1115

\bibitem[]{} Sari, R., Esin, A. A., 2001, ApJ, 548, 787

\bibitem[]{} Sari, R., Piran, T., 1995, ApJ, 455, L143

\bibitem[]{} Sironi, L., Goodman, 2007, ApJ, 671, 1858

\bibitem[]{} Sironi, L., Spitkovski, A., 2011, ApJ, 726, 75

\bibitem[]{} Spitkovsky, A., 2008, ApJ 673, L39 

\bibitem[]{} Uhm, L., Beloborodov, A., 2007, ApJ, 665, L93

\bibitem[]{} Yonetoku, D. Tanabe, S., Murakami, T. et al., 2008, PASJ,
  60, 352

\bibitem[]{} Zhang, B.-B., Liang, E.-W., Zhang, B., 2007, ApJ, 666,
  1002

\end{thebibliography}
\end{document}